# Collaboration between Research Institutions and University Sector Using Cloud-based Environment

Yuliya Nosenko[1], Mariya Shyshkina[1], Vasyl Oleksiuk[2]

[1] Institute of Information Technologies and Learning Tools of NAES of Ukraine,
9 M.Berlynskoho St., Kyiv, Ukraine

{nosenko, shyshkina}@iitlt.gov.ua

[2] Volodymyr Gnatiuk National Pedagogical University of Ternopil,
2 Maxyma Kryvonosa St., Ternopil, Ukraine

oleksyuk@fizmat.tnpu.edu.ua

**Abstract.** The high-tech cloud-based university environment formation, which would provide collaboration between the research institutions and the university sector contributes to the experience exchange; organization of joint events on sectorial, national, and international levels; improvement of the research results' implementation. It would unite the learning resources, the tools of scientific exploration support, and cover different levels of training, that could promote overcoming the gap between the research process, and the implementation and use of its results. This approach is to contribute to the development of university autonomy, which is the aim of the current reformations in higher education in Ukraine. The article represents the main results of the joint research laboratories activity that is held between the Institute of Information Technologies and Learning Tools of NAES of Ukraine and several universities being organized via cloud technology usage and devoted to its development.

## 1 Introduction

During 15years the Institute of Information technologies and Learning Tools of the National Academy of Educational Sciences of Ukraine (IITLT of NAES of Ukraine) carries out research work aimed at solving the problems of development, implementation and use of new learning tools and information technologies in education construction and development of computer-based learning environment of open education and e-learning, electronic educational resources, management and support of the researches, exploration of cloud technology. A number of important the oretical and applied results were achieved in recent years.

The relevance and the urgent need for these studies is caused mostly by the need to implement a modern educational paradigm, which consists of providing equal access to high-quality education for everyone who needs to learn who has desire and realizes need for life long learning.

Theoretical results and practical orientation of the researches conducted at the Institute are mainly subordinated to the specified educational paradigm, aimed at the development of scientific and methodological foundation of implementing the principles of open education. According to numerous studies, the implementation of the principles of open education is the backbone of the formation of global education systems, development and upgrading of learning and research environments, perspective way for national education system development.

Ukraine has achieved significant results in the study of theoretical and methodological basis of modeling and designing of information environment of open education (V. Bykov, G. Kravtsov, V. Kukharenko, A. Manako, L. Panchenko, S. Semerikov, A. Spivakovskiy, M. Zhaldak, and others). In particular, in V. Bykov's significant work the models of organizational systems of open education are designed; the models of information educational environment are proposed (methodical systems of e-learning, model of educational management in its different organizational levels, modern informatics teachers' training in higher educational institutions and others) [6]. The named works are aimed to create the methodological basis for further researches in this area, given that the cloud oriented environment is a new step in the development of open education.

General trends of implementation of promising ICT, including cloud technology, into organization of educational systems were studied in the works of V. Bykov, V. Kumar, S. Litvinova, N. Morze, V. Oleksyuk, S. Semerikov, Z. Seydametova, M. Shyshkina, A. Stryuk, M. Zhaldak, Yu. Zaporozhchenko (Nosenko) [15, 16], and others.

Problems of creation of cloud oriented learning environment in higher educational institution were studied in works of V. Bykov, A. Fox, R. Griffith, V. Kukharenko, V. Oleksyuk, L. Panchenko, S. Semerikov, M. Shyshkina, K. Slovak, A. Spivakovskiy, A. Stryuk, K. Subramanian,N. Sultan, Yu. Tryus, M. Zhaldak, and others.

In Z. Seydametova's paper [21] the analysis of the use of cloud services in higher school for educational and scientific purposes, and their systematization is proposed.

The main types of service models that reflect the possible ways of ICT-outsourcing use for software and computing resources access provision are considered in M. Shyshkina's paper [23].

The issues of determining the most appropriate technological solutions for corporate network of the cloud-based learning environment of educational institutions are examined by Ukrainian and foreign researchers [24].

S. Semerikov in his paper [12] analyses the term "cloud services" and its evolution, which outlines the prospects and trends of these services and their application to support the processes of research activities. The use of cloud services in maintaining the processes of research activities is investigated by O. Spivakovsky [32], O. Spirin [31], A. Stryuk [33].

The study of A. Alkhansa, A.A. Shakeabubakor, E. Sundararajan, and A. Hamdan [1] is devoted to issues of productivity of university research processes, and it is proposed to use cloud services as a tool to increase efficiency and enhance the research activities.

Indeed, the cloud technology implementation will provide each educational institution with the following advantages and capabilities: to use the modern, systematically renewed IT infrastructure, software and services at no additional cost; reduce the cost of development and support of local information networks; engage the education participants' personal computer devices into the learning process; ensure the consolidation of electronic educational resources in a single electronic repository; implement mobility of educational process participants, support the distant, blended learning and other innovative forms and technologies [1], [7], [22], [27], [28].

Currently, the potential of cloud technology is engaged not enough in the practice of Ukrainian education. There is a growing discrepancy between the level of our local schools IT-infrastructure and modern European ones. These contradictions might cause a significant technological gap in ICT-equipment of higher educational institutions and even lead to a significant gap between the levels of their equipment in comparison with the similar ICT structures in foreign universities.

Therefore, an important task is to develop the learning and research environment of educational institutions in Ukraine considering the latest developments in the field of scientific and technical progress.

One of the possible ways for this implementation is the organized cooperation among research institutions and university sector, providing the joint research laboratories within the cooperation agreements to implement the results of scientific studies, involving representatives of educational institutions to scientific researches, improvement of teaching staff training, encouragement of participation in the development of joint projects on sectorial, national, and international levels.

The **purpose** of the article is to describe the state of the art of the cloud-based learning and research environment formation in Ukrainian university sector and to expose the research results on its deployment within collaboration between the research institution and the university sector.

## 2 The Presentation of Main Results

### 2.1 The Activity of the Institute of Information Technologies and Learning Tools of NAES of Ukraine in Education ICT Support

The National Academy of Educational Sciences of Ukraine (NAES of Ukraine) is the main organization on the national level which provides methodological and scientific support, legal and normative basics for the development of concepts and approaches to education improvement on different levels.

The Institute of Information Technologies and Learning Tools of NAES of Ukraine (http://iitlt.gov.ua/) is the principal research institution on scientific and methodological support of formation and development of electronic educational space in the field of general, special, vocational and teachers training and adult education and also on ICT support of NAES of Ukraine. The Institute carries out computer and technological support of electronic resources of the research and educational institutions within NAES structure, and provides representation of these institutions via Internet [3].

Now the computerization covers all the basic functions of NAES, such as scientific, educational, administrative, informational. The corporate automated information system of NAES of Ukraine is developed systematically. It is connected to URAN, and therefore to GEANT. We should mention that NAES of Ukraine is its founding member since 2006 (http://www.uran.net.ua/~ukr/uran-statut.htm). And the Institute of Information Technologies and Learning Tools is a member of the Association of users of Ukrainian research and Academic network "URAN" (URAN - Ukrainian Research and Academic Network) since April 12, 2007 (http://www.uran.net.ua/~ukr/uran-members.htm).

The corporate automotive and information system of NAES forms a single computer and technological foundation of information support of NAES, and its principal components are: scientific projects; computer technologies; information, personnel, organizational, managerial, financial and economic resources. Tools and technologies that support corporate networks and electronic information resources of NAES such as sites of its scientific and educational institutions, various computerized databases, information portals, e-learning courses, etc. provide automated systems of research, electronic communications in the Internet space and maintain information process of creating electronic textbooks, manuals and other scientific and educational electronic editions [4], [5], [7].

There is a range of information resources and systems developed at the Institute: the official website of IITLN of NAES of Ukraine (http://iitlt.gov.ua/), a peer-reviewed journal in educational sphere, which publishes full-text article online bimonthly with immediate open-access (http://journal.iitta.gov.ua), Digital Library of NAES of Ukraine (http://lib.iitta.gov.ua/), conference supporting system “Educonference” (http://conf.iitlt.gov.ua/), and others.

We can trace trends in the implementation and use of web-oriented open corporate information systems in NAES of Ukraine through the activities of IITLT of NAES of Ukraine during recent years. Thus, the electronic scientific specialized journal “Information Technologies and Learning Tools” (http://journal.iitta.gov.ua/index.php/itlt/index) was launched in December 2006, added to the list of electronic scientific professional editions in pedagogical sciences of Ukraine in April 2007. It is bimonthly open-access peer-reviewed e-journal that includes full-text articles in theory, methods and practice of using ICT in education. Since 2011 the edition began functioning on Open Journal Systems (OJS), developed by the Public Knowledge Project (Canada), which is an open information-technological platform for the deployment of scientific journals. Since 2012 the journal is indexed in scientometric databases, including such as Google Scholar, Index-Copernicus, Universal Impact Factor (UIF) et al., and numerous abstracts databases (the Directory of Open Access Journals (Sweden), Academic Journals database (Switzerland), National abstracts database “Ukrainikanaukova” (Ukraine) and many others). Currently the journal holds the 8th place according to “The rating of scientific journals of Ukraine”, compiled by The Vernadsky National Library of Ukraine. According to Google Scholar as of April 1, 2016 the number of journal’s articles citations is 2615, h-index – 21, i10-index – 82 [8].

The Digital Library NAES of Ukraine (http://lib.iitta.gov.ua/) is another significant project of the Institute, and it was founded during 2009-2011, developed on the EPrints platform. Using statistical module IRStatsit is possible to obtain the data on the quantity and quality indicators over the Library for downloads of all scientific productions or separate publications distributed within the specific collections (subdivisions by scientific institutions and their divisions, author name, period of time, topic of research, etc.) [8].

Since 2012 the web-translation tools for distant participation and conducting different scientific events (conferences, workshops, etc.) are used, particularly via Skype, Wiziq platform and others.

In 2014 the institutional system for conference support “Edu-conference” was projected and implemented[30].By means of this platform the support of submission and reviewing the materials of participants of Ukrainian Scientific Conference for Young Scientists "NaukovaMolod’" and annual reports for the conferences of IITLT of NAES of Ukraine is held.

Development of experimental activity of IITLN of NAES of Ukraine, which provides practical implementing of scientific studies into the educational process, is carried out within joint research laboratories, research projects at the regional and national levels, pedagogical experiments, etc.[22],[24,25,26].

### 2.2 The Expediency of Cooperative Work between a Research Institution and University Sector

Formation of a high-tech learning environment based on cloud technology, which would unite educational resources for learning purposes, support of scientific research, and cover different learning levels, could promote overcoming the gap between the research process, and the implementation and use of its results(that is a significant problem in context of Ukrainian education and science interaction) [27].

One of the possible ways to develop learning and research environment of educational institutions of Ukraine considering the latest developments in the field of scientific and technical progress is the cooperation among the research institutions and university sectorial, providing the joint research laboratories to implement the results of scientific studies, involving representatives of educational institutions to scientific researches, improvement of teaching staff training, encouragement of participation in the development of joint projects on sectorial, national, and international levels [27].

Establishment of joint research laboratories contributes to the spreading of the experience of activity management and its results; organization of joint events on sectorial, national, and international levels; improvement of the organizational techniques of research results’ implementation; overcoming the potential gaps between conducting scientific researches and the irpractical use.

This approach can contribute to the development of university autonomy, which is the aim of current reformations in higher education and is consistent with the adoption of the new Ukrainian Law on Higher Education.

The research areas of scientific and educational institutions are coordinated by developing the links with the joint laboratories, and thus the focus is made at those pedagogical issues, being currently at most need in educational space. By virtue of this the mechanisms of the research results’ implementation are improved. The proc-esses of teaching staff training occur in close cooperation between researchers and teachers, in interrelation of educational process and scientific research.

On the basis of modern network technologies the possibility to appeal to the remote educational resources online appears. For example, it can be realized with the use of virtual laboratories and laboratory systems of remote access, university resource rooms and laboratories for conduction of demonstration experiments [27].

Trends associated with the processes of integration of universities educational environments, suggest their participation in regional clusters formation. Clusters are a form of cooperation in the field of scientific, research and innovative activity and are formed through the merger of companies and organizations relevant to a particular kind of industrial activity [10]. Cooperation can take the form of information exchange, resource sharing, pooling in terms of staff training and employment. In particular, one of the advantages of university clusters is the transfer of non-core functions of organizing and maintaining of the university ICT infrastructure functioning to the experts. To realize this purpose a separate IT department is created in the cluster [10], [27]. Thus the functioning of the high-tech infrastructure is conducted from a single center through outsourcing, i.e. ICT services required by a system are implemented through another external system.

The trend towards consolidation of higher educational institutions has gained acceptance abroad [13], [19], and is manifested in Ukraine. It consists of creating regional universities, which may contain several higher educational institutions. Implementing shared technological platform of the regional educational institution functioning on the basis of cloud computing is a way to solve a number of problems that arise when combining infrastructural ICT into a unite framework. In its turn it allows to access to the best examples of electronic educational tools and resources by those institutions that don't possess the necessary financial support and powerful ICT units [27].

Besides, within the interaction network the university cooperation with research institutions and business structures can be realized, the processes of training, skills development, international projects implementation, linkages between schools and universities can be conducted.

This is consistent with the prospect of creation of integrated (sectorial, national) databases, data collections, resources that are made available to various educational institutions [9], [25]. In order to take advantage of such collections in full, it is also useful to implement tools of cloud computing [27].

### 2.3 The Joint Laboratory of the Institute of Information Technologies and Learning Tools of NAES of Ukraine, and Kryvyi Rih National University

For the aim of coordination and further development of the researches the Joint laboratory "Cloud Computing in Education" (CCELab) was created on the basis of Kryvyi Rih National University and IITLT of NAES of Ukraine (2012).

The main goal of the CCELab is methodological and experimental research of emerging e-learning technologies and exploration of different aspects of cloud computing application for education and personnel training.

So, main objectives of the CCELab are:

- coordination of research and development on the problems of using cloud technology in education, carried out at IITLT of NAES of Ukraine, National University of Kryvyi Rig, and some other national universities and institutions;
- development, testing, implementation and experimental approve of cloud cloud-based e-learning platforms, tools and services;
- investigations on cloud-based learning environments' modeling and application;
- experimental study of cloud-based e-learning infrastructures for education and training of professionals;
- research for fundamental, methodical, technological, organizational and other use features, presupposition of an introduction and perspective ways of using cloud technology in education;
- publication and discuss of results of scientific and experimental study of the laboratory;
- participation in the organization of international and national scholarly conferences, workshops, etc.

Within the joint activities a range of electronic resources was developed: the site of joint research laboratory (http://cc.ktu.edu.ua/); scientific and educational cloud of the Department of Cloud Oriented Systems of Education Informatization of IITLN of NAES of Ukraine, which is used to organize joint activities and resources exchange; cloud oriented educational and scientific laboratory environment. A unique experience is the use of OwnCloud – software with open source, designed for development, modification of specific applications required for the organization of scientific and methodological studies.

In addition to the above, an annual international seminar "Cloud technologies in education" was founded within joint laboratory activity. The seminar is attended by representatives of more than 50 educational institutions from more than 20 cities of Ukraine and other countries – Russian Federation, Australia, USA, etc. The proceedings of the seminar are available in open access (http://cc.ktu.edu.ua/report.html). Also, the annual surveys on the development and use of cloud computing services in schools are held.

To show the state of the art of the cloud-based learning environment development and the rate of cloud-based services use by educational personnel in Ukraine the survey was made within the framework of the International internet-seminar "Cloud Technologies in Education" that is held annually since December, 2012. More than 120 members of 54 educational institutions from 22 cities and 18 regions of Ukraine took part in this seminar. The participants were concerned with the problems of cloud technology, well acquainted with the modern trends of technological development, and their organizations were well equipped and oriented for using advanced ICT. So, we involved them into appropriate interrogation.

The responses to the question: "What type of the cloud-based services is mainly used in your educational institution?" showed that 78% of participants had chose SaaS (Software-as-a-Servise), 29,3% – PaaS (Platform-as-a-Service), 22% – IaaS (Infrastructure-as-a-Service); and 31,7% – DaaS (Desktop-as-a-Service).The results are presented in Fig. 1 (the entries are not excluded).

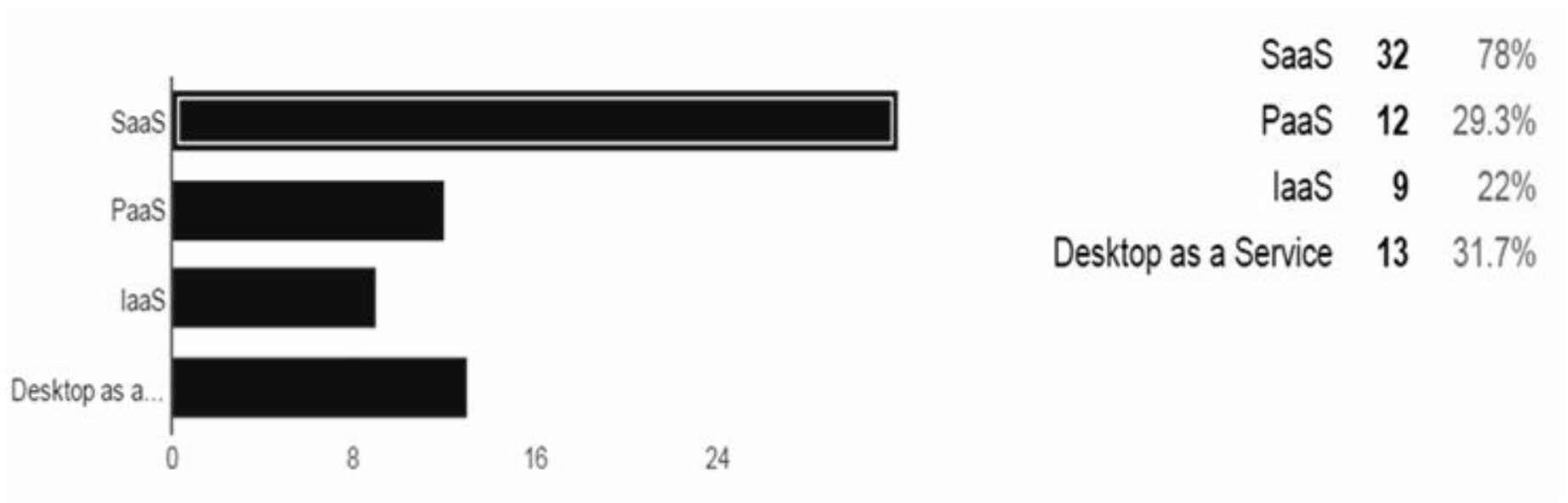

**Fig.1.** The results of the survey on cloud technology use at the institutions of higher education in Ukraine (2015)

To the question: "What type of the cloud-based platform or services are used in you institution?" we got the next results: Google Apps for Education – 76%;Microsoft Office 365 – 51,1%; Special Software (for example, the SageMathCloud) – 17%; 10,6% use public cloud services, such as Amazon, Microsoft Azure, etc.); 6,4% use services of the private cloud (Microsoft Azure, Xen, WMWare, etc.). These results are presented at Fig. 2 (the entries are not excluding).

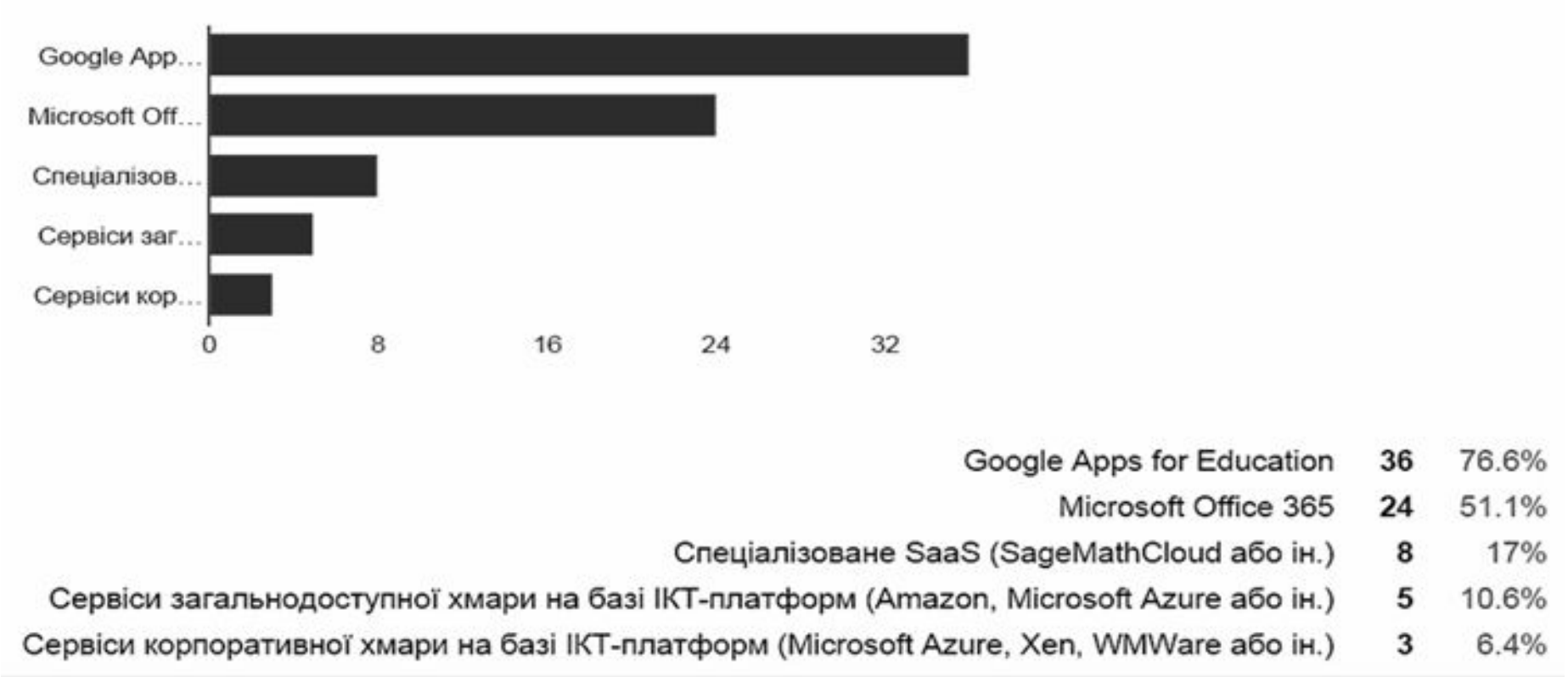

**Fig. 2.** Application of cloud-based platforms and services at the educational institutions in Ukraine (2015)

As it appears from the study, the cloud-based services are widely used in Ukrainian educational institutions. But still their use is to be more systematic, to be organized into united system, to be consciously and purposely oriented at pedagogical aims. So there is a current need for the upgrading of ICT competence of educational personnel, mainly those ones who engage in providing educational systems with emerging ICT, in particular, public administration employees [28].

## 2.4 The Joint Laboratory of the Institute of Information Technologies and Learning Tools of NAES of Ukraine and Kherson State University

In 2011 our joint research laboratory on issues of using ICT in control of learning quality was created at Kherson State University. In 2013 the University was approved as the experimental base for research work of the regional level: "Computer-oriented quality management system of electronic educational resources in schools" (2013-2015). The experiment involved four pilot schools from Kherson region.

The aim of the experiment is to determine and verify experimentally the requirements and methods of evaluating the quality of electronic educational resources (EER) in pilot schools learning process. As a result of the experiment the didactic requirements and methodology for evaluating the quality of EER were determined and tested [26]. The significant part of the experiment was the development of the performance indicators for evaluating the quality of cloud oriented components and methods forevaluating the quality of electronic learning resources in the cloudoriented environment. In 2014 within the experiment the cloud storage of open informational resources via Google Drive was created: https://drive.google.com/folderview?id=0BwItlyWdtLeIbU5pUWR6c0NiWTQ&usp=sharing. Thus, the experts have opportunity of sharable documents processing, editing, and updating, relating to expert evaluation activity.

The evaluation of the quality of EER in the cloud-based learning environment is a separate trend of the Laboratory activity. In this case there are different approaches and indicators. The access organization has been changed, so the models of learning activity have been changed also. That is a problem: what features and properties have to be checked to measure the pedagogical effect of the cloud-based approach? How to assess the pedagogical innovation, what are the factors of its influence for pedagogical systems, its structure and organization? Were the learning results improved due to the cloud-based models? In this context just the quality of EER is a criterion to estimate the level of organization and functioning of the cloud-based learning environment [22].

The perspective way of estimating the quality of EER is to be made by means of the cloud-based environment. As the resources are in collective access so there is a way to involve experts into the learning process so they could observe and research the functioning. This is a way to make the process of quality estimation more convenient, flexible and quick. The process of estimation becomes anticipatory and timely. The estimation may be obtained at ones with the process of EER elaboration, and this is very important to facilitate the process [22], [26].

The cloud-based learning component used in the experiment has undergone the quality estimation. The method of estimating the quality of learning resources developed in the joint laboratory of educational quality management via ICT [11] was used and adapted for this study.

The advantage of the approach is in possibility to compare the different ways of resources implementation with regard to the learning infrastructure set. There is a perspective of future research in this direction as for taking into consideration different types of resources and environments.

## 2.5 The Joint Laboratory of the Institute of Information Technologies and Learning Tools of NAES of Ukraine and Ternopil Volodymyr Hnatiuk National Pedagogical University

The joint laboratory was created for consolidating and coordinating the research work of scientists of IITLT of NAES of Ukraine and the Department of Informatics and methods of its teaching of Ternopil Volodymyr Hnatiuk National Pedagogical University. The general purpose of laboratory is implementation of cloud technology in the educational process of higher and secondary education institutions.

The objectives of lab researches are:

- integration of cloud-based and traditional applications into IT-infrastructure of universities;
- implementation and use of public clouds for learning process support;
- analyzing, testing, deployment, implementation and experimental approve of the private cloud platforms in higher education institutions;
- creating virtual laboratories based on corporate clouds, and their technical and pedagogical support;
- development of methodology of virtual labs application in the learning process.

We consider clouds as tools for solving the learning problems. Therefore, researchers help in deploying school clouds based on public platforms and obtaining academic licenses for them.

Scientists of joint laboratory researched the integration of cloud services Google Apps into information and educational space of higher or senior educational institution. They have offered that integration has provide unification authentication and access to learning objects. The unified authentication was designed on the base LDAP-directory. We have adjusted programs for synchronization of the LDAP-directory and public clouds (Google Apps, Microsoft Offlice 365) [17].

The participants of laboratory investigated models deployment of cloud technology in IT-infrastructure. The hybrid model have recognized most suitable to IT-infrastructure of higher educational institution. Open source platform for organizing enterprise clouds analyzed and tested during last year. We have identified IaaS model as most functional for enterprise clouds. The model of private academic cloud was developed.

The enterprise cloud of Department of Physics and Mathematics of Ternopil V. Hnatyuk National Pedagogical University deployed (http://cloud.fizmat.tnpu.edu.ua – available only from university network or through VPN). We have used Apache CloudStack as platform for this cloud [18]. Deployed cloud infrastructure contains such elements: 1 zone, 1 pod, 1 cluster, 3 hosts, 3 primary storages and 1 secondary storage.

In our cloud, Apache CloudStack provides:

- run above 100 virtual machines (VMs);
- connect VMs through simple and advanced networks;
- access to VMs through web-interface and standard protocol such as RDP and SSH;
- distribution of computing resources for VMs;
- create template and snapshot of VMs;

- authentication through LDAP protocol.

We had deployed in cloud infrastructure several virtual networks. We can associate it with physical networks of IT-infrastructure. This allows to model routing process in Internet.

Scientists of joint laboratory are developing virtual laboratories for learning computer sciences, it is described in detail in the article[22].

Three joint research laboratories mentioned in the article were chosen to be the experimental base for the implementation of research on the topic: "Methodology of formation of a cloud-based learning and research environment of educational institution" (Decision of the Scientific Council of the Institute of Information Technologies and Learning Tools of NAES of Ukraine, №4 of March 04, 2015).

Within this research the overall investigation of the ICT competence of the students and lecturers was undertaken. In the focus of investigation there was research and educational community which was formed around the cloud-based trend of research. There were those engaged in research and training actions, seminars, webinars, conferences other events conducted by the joint laboratories involving the vast number of educational institutions. Besides three institutions were joint laboratories were established other institutions where collaboration was carried out under agreements on cooperation with IITLT of NAES of Ukraine also were included.

Among them there were:

- National Pedagogical University, named after M.P. Dragomanov;
- Drohobych State Pedagogical University, named after I. Franko;
- Cherkasy State Technological University.

While forming the control and experimental groups of the experiment the next conditions were taken into consideration:

- In the control group there were those participants of the experiment who had taken part in scientific and methodological events of the Institute and had shown significant interest in the use of cloud technology in their research and educational activity; but they had not be engaged into special training of research projects of the Institute, devoted to the cloud computing tools application – lecturers, researchers, post graduate students and others (58 participants);
- In the experimental group there were those members who also had belonged to the scientific and educational community concerned with the cloud computing applications still they had been members of joint research laboratories and had taken part in special training devoted to CC application and also had been involved into research work in which such training had been supposed (60 participants).

Analysis of the results of pedagogical experiment formative stage showed that the distribution of ICT competencies in experimental and control groups is characterized by statistically significant differences due to the use of cloud services for educational and scientific purposes and also use of the appropriate methods for their application in the process of scientific and pedagogical staff training (Fig. 3).

Another stage of the experiment was related to a survey of students. The experimental group consisted of students who had studied math or informatics with the use of cloud services due to the specially developed technique (overall by all experimental base – 50 students).

The control group consisted of students who had studied the same subjects and had been familiar with cloud services but used it without a specially developed method (120 students). Survey results are showed in Fig. 4. The distribution of ICT competences had statistically significant differences due to the use of the special technique.

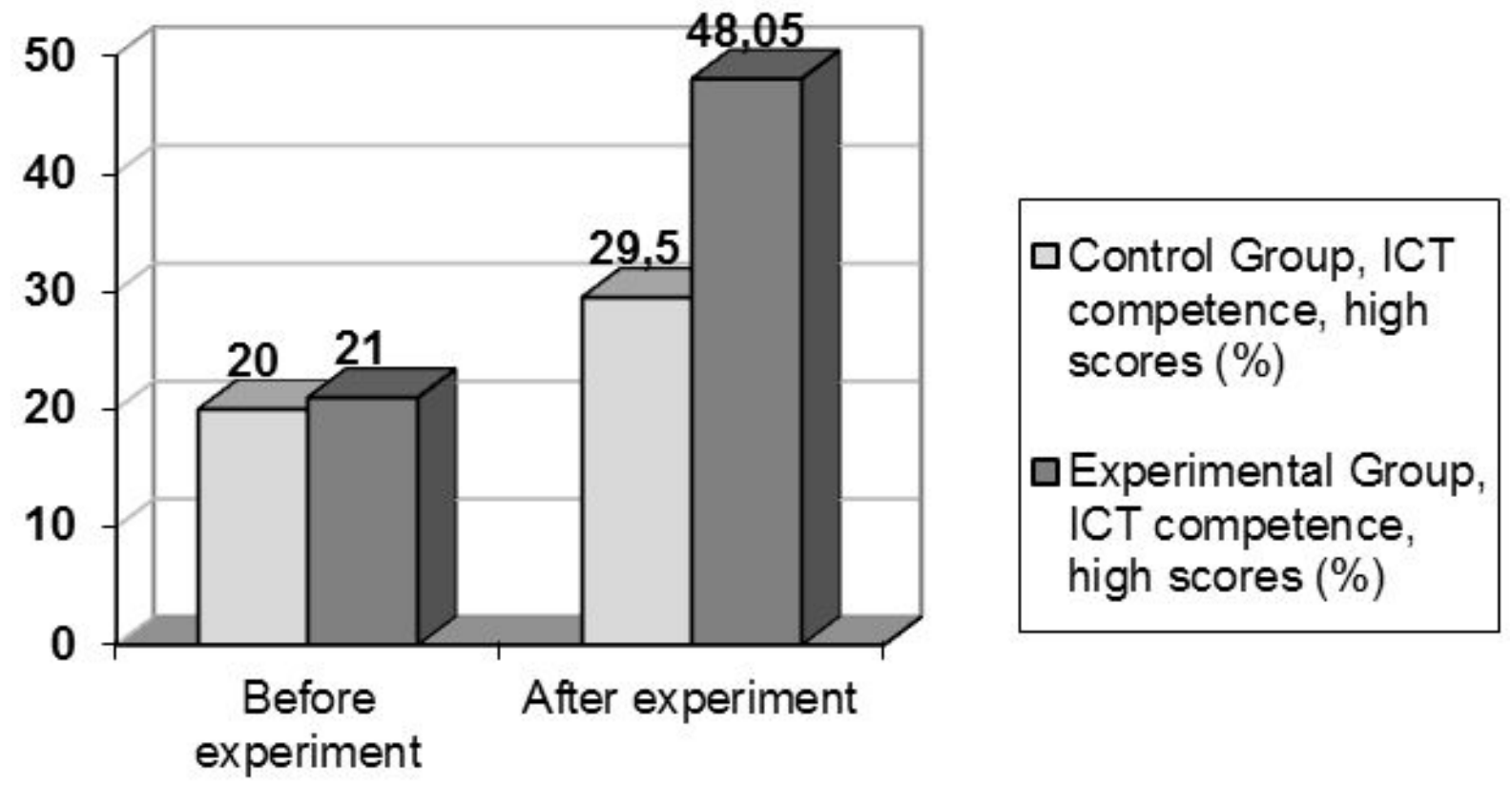

**Fig. 3.** Measuring the level of lecturers' ICT competence before and after the experiment

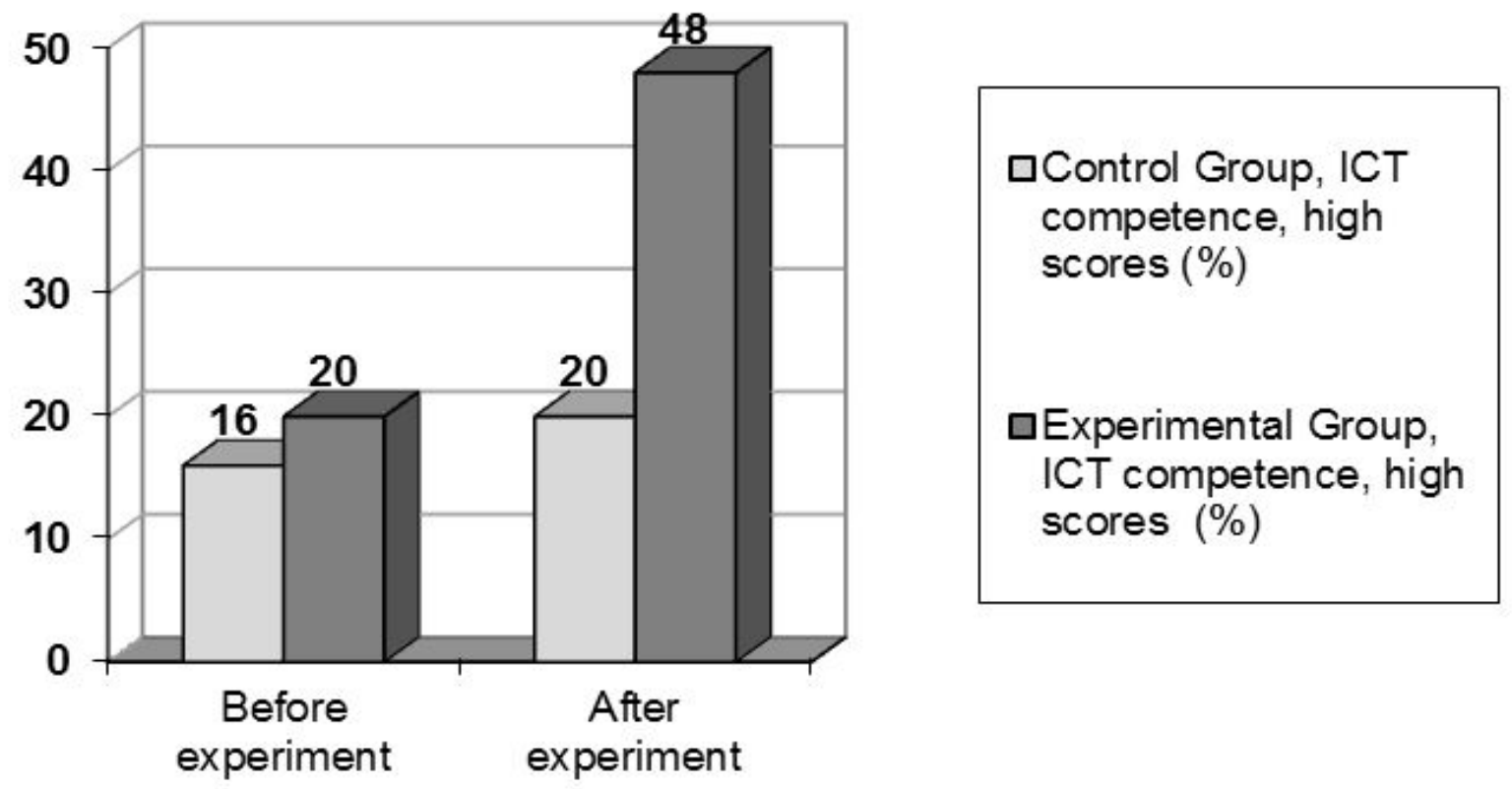

**Fig. 4.** Measuring the level of students' ICT competence before and after the experiment

Thus, the analysis of the pedagogical experiment results showed that there had been a growth of ICT competencies of students and pedagogical stuff of educational institutions, that had had statistically significant differences, indicating the efficiency of the cloud-based learning and research environment formation and relevance of the special technique that was used in the training of scientific and pedagogical staff.

## 3 Conclusions

Within the research the most important scientific and methodological problems which concern the development of learning and research environment of educational institutions in Ukraine were revealed. Among them there are: the pedagogical and technological parameters of the learning environment modernization determining; the methods of research and pedagogical personnel training development with the use of the cloud-based tools; the learning content improvement due to the cloud computing systems application for different learning disciplines; elaboration of methodical guides for the use of the cloud-based learning tools for IT-disciplines.

The specific complex of problems is connected to the use of cloud computing in the field of inclusive education which is one of research priorities of European scientific and educational community as it is reflected in Horison 2020 international program for research and innovations.

Among the possible ways of improving learning and researh environment of higher educational institutions there are:

- wider use of emerging ICT tools and network technologies in the scientific and educational practice as its capacities essentially increased due to the use of cloud technology;
- more active involvement of international network information systems and infrastructures into scientific and educational activity of universities, so as to promote more openness and integration of scientific research and applications;
- organization of international cooperation, coordination and harmonization of national and international strategies for scientific and technological development, broader integration into the European educational and research space;
- expansion of cooperation among educational and research institutions; establishment of corporate structures of various nature (affiliated schools in manufacturing, training and engineering centers, research laboratories etc.) aimed at the development and implementation of advanced ICTs in various spheres of social activity, in particular in education;
- defining the prioritized and relevant researches in IT areas, particularly concerning educational technology and software for learning purposes to attract the attention of society, public sector and industry to this issues.

A promising solution for modernization of the learning and research environment of educational institutions is the development of cooperation through the organization of joint research laboratories that may contribute to the spreading the research results; organization of joint events on sectorial, national, and international levels; improvement of the organizational techniques of research results' implementation; clearing the research priorities; overcoming the potential gaps between conducting scientific researches and their practical use, and so on.